# Understanding Transfer Learning for Chest Radiograph Clinical Report Generation with Modified Transformer Architectures

Edward Vendrow*
*Department of Computer Science*
Stanford University
evendrow@stanford.edu
*Contributed Equally

Ethan Schonfeld*
*Department of Biomedical Data Science*
Stanford Medicine
eschon22@stanford.edu
*Contributed Equally

*Abstract*—The image captioning task is increasingly prevalent in artificial intelligence applications for medicine. One important application is clinical report generation from chest radiographs. The clinical writing of unstructured reports is time consuming and error-prone. An automated system would improve standardization, error reduction, time consumption, and medical accessibility. In this paper we demonstrate the importance of domain specific pre-training and propose a modified transformer architecture for the medical image captioning task. To accomplish this, we train a series of modified transformers to generate clinical reports from chest radiograph image input. These modified transformers include: a meshed-memory augmented transformer architecture with visual extractor using ImageNet pre-trained weights, a meshed-memory augmented transformer architecture with visual extractor using CheXpert pre-trained weights, and a meshed-memory augmented transformer whose encoder is passed the concatenated embeddings using both ImageNet pre-trained weights and CheXpert pre-trained weights. We use BLEU(1-4), ROUGE-L, CIDEr, and the clinical CheXbert F1 scores to validate our models and demonstrate competitive scores with state of the art models. We provide evidence that ImageNet pre-training is ill-suited for the medical image captioning task, especially for less frequent conditions (eg: enlarged cardiomediastinum, lung lesion, pneumothorax). Furthermore, we demonstrate that the double feature model improves performance for specific medical conditions (edema, consolidation, pneumothorax, support devices) and overall CheXbert F1 score, and should be further developed in future work. Such a double feature model, including both ImageNet pre-training as well as domain specific pre-training, could be used in a wide range of image captioning models in medicine.

Keywords: pre-training, transfer learning, memory-meshed transformer, cheXpert, radiograph, natural language generation

## I. Introduction

In the medical domain, a task that appears in almost all specialities is the generation of reports from medical imaging. Whether this imaging is simple 2D chest radiographs or 3D time series of functional brain activity mappings, experienced clinicians generating many such reports daily are error prone. In the medical setting, such errors could prove fatal. Advances in deep learning based image captioning allow for the potential automation of such clinical tasks.

Corresponding Author

The release of the MIMIC-CXR dataset [1] inspired multiple efforts for chest radiograph image captioning. The field existed prior to the 2019 release; however, it was limited to datasets using only a few thousand matched radiograph to report examples. With the release of MIMIC-CXR containing 371,920 chest radiographs with 227,943 imaging reports, more advanced models making use of novel transformer architectures catalyzed performance in the growing field.

Deep convolutional neural networks [2] have revolutionized the field of computer vision, in part due to the discovery that supervised pre-training for an auxiliary task, followed by fine-tuning on the desired task, significantly improves performance [3, 4]. This process generally involves training on a large-scale dataset such as ImageNet [5] followed by a target task with less training data. Further attempts to improve performance with pre-training have used even more data, up to 3000x the size of ImageNet [6, 7]. However, recent research has challenged this conventional wisdom of 'pre-training and fine-tuning' in computer vision: He et al. [8] reported that ImageNet pre-training does not improve performance on the COCO object detection and instance segmentation tasks compared to random initialization, given enough training iterations to properly converge. Mathis et al. [9] similarly showed that even for small datasets, while ImageNet pre-training is helpful for in-domain tasks, it does not necessarily improve out-of-domain generalization to unrelated tasks.

Most medical image captioning models use ImageNet pretrained models as radiograph image feature extractors, sometimes without fine-tuning, or don't use pre-training at all [10, 11, 12, 13, 14]. However, recent work has demonstrated that ImageNet pre-training may not give much information for medical applications. ImageNet was further shown to apply to medical classification tasks, but not for segmentation tasks due to largely homogenous imaging characteristics with limited morphological information [15]. Recent work, focused on the domain gap between natural images and medical images, has even attempted to develop unsupervised pre-training strategies for radiography applications to substitute

ImageNet pretraining for domain specific methods [16]. Thus, it becomes crucial to determine the suitability of ImageNet pre-training for the image captioning task of clinical report generation from chest radiographs.

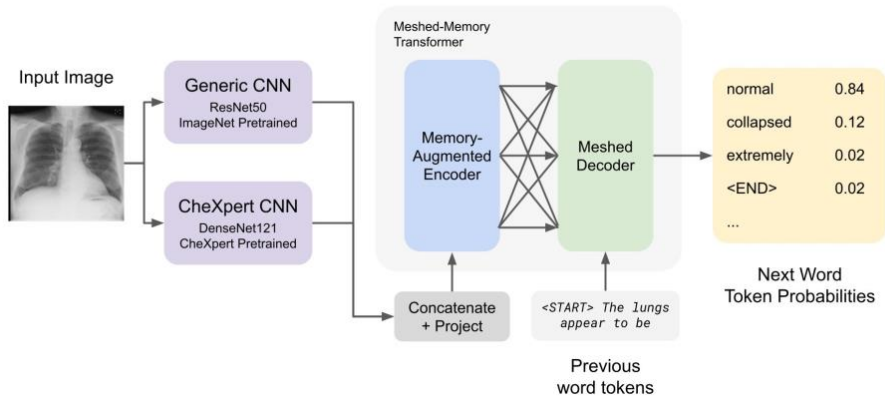

Fig. 1. Baseline model architecture. The outputs of both CNNs are concatenated and projected into the encoder and then into the decoder to generate the final text output.

A significant advantage of studying the appropriateness of ImageNet pretraining in the domain of chest radiographs is the existence of CheXpert, a large dataset of labelled chest radiographs [17]. The Stanford Machine Learning Group directed competition on this task resulted in development of state of the art CNN based models, achieving AUC of 0.930, to predict the presence of 14 observations from chest radiographic image inputs [18]. Such highly domain specific models trained on CheXpert have only been used by a few groups working to generate radiology reports [19]. Here we propose a systematic investigation of model performance, especially related to its clinical success on each of the 14 radiological observations, by whether its feature extractor is pretrained on ImageNet, CheXpert, or both (using a novel double feature architecture).

II. RELATED WORKS

A. Image to Text Radiology Clinical Report Generation:

Past approaches to clinical report generation include LSTM and other RNN methods, but the most promising recent works in this domain use the transformer model [20]. Chen et al. [11] use a meshed-memory transformer, which has shown promise in image captioning tasks by using a multi-level representation of the relationships between image regions. Xiong et al. [21] further apply reinforcement learning to the output text generation to improve generated report quality, treating next word selection as a task performed by an RL agent. Their contribution is notable because word selection is a discrete task and thus non-differentiable without using other tricks, so using an RL agent bypasses this restriction. Liu et al. [22] extend this work to add a further RL reward for clinical coherence to improve medical relevance of the generated reports. Meanwhile, Syeda-Mahmood et al. [23] simply use existing reports as templates to generate new ones. Significantly, Miura et al use the Meshed-Memory transformer as their base architecture and supplement this with two rewards in a reinforcement learning system, demonstrating that optimizing traditional NLG metrics does not maximize clinical F1 success. Following state of the art work from Chen et al and Miura et al, we use the meshedmemory transformer for our architecture in this analysis.

B. Imagenet Pre-training versus Domain Specific Visual Extractors:

While pre-training on ImageNet was found to provide a significant boost in performance for chest radiograph interpretation, ImageNet was found to provide only a small boost for larger model architectures such as those for the medical image captioning task [24]. In the medical domain, Raghu et al. [25] find that ImageNet pre-training does not significantly benefit performance on medical imaging tasks, especially as compared to simple, lightweight models. Surprisingly, Kornblith et al. [26] show that the performance of pre-trained models on ImageNet (implicitly used as a predictor of how well a model will transfer-learn) can actually correlate negatively with performance on other vision tasks. These results suggest that not only is out-of-domain pre-training on ImageNet often unhelpful, but pre-training may also give a misleading intuition on the performance of transfer learning.

III. METHODS

## A. Double Feature Transformer

We perform clinical note generation via a modified transformer architecture. Rather than using a single CNN backbone to extract image features, we use two: an ImageNet pretrained encoder, as is standard in the literature, and a chest radiograph specific CNN trained on CheXpert labels (ranked 5th in CheXpert leaderboard, AUC=0.929) [27]. The CNN backbones yield a grid of feature vectors (e.g. 8x8x1024), which are then projected to the embedding dimension of the transformer. For the double feature model, we use two separate CNN backbones to encode the image features. The respective outputs are concatenated and a linear layer applied to reduce the dimensionality before feeding them to the encoder for further processing. As is standard, the transformer decoder outputs a discrete softmax probability over the vocabulary predicting the next word. NLL loss is used. This model architecture is described in Figure 1. In order to further improve performance, we also use a meshed-memory transformer as used by Chen et al. [11]. Since the meshed-memory transformer architecture is similar to a regular transformer, no additional modifications are needed to adopt a double-feature architecture beyond the changes previously described.

## B. Training Details

We train 3 models: 2 single-encoder models using ImageNet and CheXpert pre-trained CNN backbones, and a doublefeature model. The ImageNet CNN architecture is ResNet50, while the Chexpert backbone architecture is DenseNet121. The transformer models use a meshed-memory transformer memory size of 40 and a hidden dimension of 512. Models are trained with a batch size of 24 for 32 epochs using the Adam optimizer. Training occurs on an AWS instance with an NVIDIA Tesla T4 graphics card.

## C. Evaluation Metrics

To evaluate the generated reports we used, we compare them to ground truth clinical reports using BLEU(1-4), ROUGEL, and CIDEr scores. These metrics have been shown to be domain agnostic and reward grammatically correct but clinically irrelavant models [28, 29]. As the field of radiology begins to move towards a structured report, a metric for correct labeling of common conditions and findings in the reports is required. To accomplish this we use the CheXbert labeller to calculate the CheXbert Clinical F1 Metric.

## D. CheXbert Clinical Metric

We use the CheXbert labeler model [30] to extract diagnosis labels from both the ground-truth and generated report, and calculate the F1 score between the two labels. The CheXbert model gives labels Positive, Negative, Blank, or Uncertain for a series of 14 medical conditions; we treat only the "Positive" label as a positive prediction for the purposes of calculating the $F_1$ score. We denote this score the CheXbert $F_1$ score.

Given ground-truth and predicted reports, we extract labels for 14 medical conditions using CheXbert, then calculate the CheXbert $F_1$ score formulated as

$$F_1 = 2 \cdot \frac{precision \cdot recall}{precision + recall} = \frac{TP}{TP + \frac{1}{2}(FP+FN)}$$

| Model | BLEU-1 | BLEU-2 | BLEU-3 | BLEU-4 | ROUGE-L | CIDEr | CXbert |
|---|---|---|---|---|---|---|---|
| ImageNet | 0.3198 | 0.1925 | 0.1232 | 0.0874 | 0.2565 | 0.3369 | 0.4195 |
| CheXpert | 0.3414 | 0.2141 | 0.1473 | 0.1084 | 0.2716 | 0.4115 | 0.5116 |
| Double Feature | 0.3476 | 0.2182 | 0.1504 | 0.1062 | 0.2728 | 0.4107 | 0.5152 |
| Chen et al. | 0.353 | 0.218 | 0.145 | 0.103 | 0.277 | – | – |

Fig. 2. BLEU 1-4, ROUGE-L, CIDEr, and CheXbert F1 metrics achieved by our models on different pre-training methods, as compared to another meshedmemory transformer method. Note that CIDEr and CheXbert F1 metrics are not reported by the authors.

Additionally, a CheXbert $F_1$ score is calculated for each of the 14 conditions separately to better inform pre-training choices based on desired medical application in future work.

## IV. RESULTS

The model results, listed in Figure 2, demonstrate that both CheXpert and the double-feature model outperform the ImageNet pre-trained model. The double-feature model achieves the highest BLUE(1-3) scores and highest ROUGE-L score, and approximately equal BLUE4 score and CIDEr score to the CheXpert model. The double-feature model achieves the highest CheXbert $F_1$ score; however, it is important to note that the differential between the CheXpert model and the ImageNet model is far greater than the differential between the double-feature model and the CheXpert model in clinical as well as standard generation metric performance. Furthermore, observe that the factor differential between double-feature model and ImageNet model performance scales increasingly with BLUE(1-3), and plateaus for BLUE4.

The metrics listed in Figure 2 validate each of the three model's performance as the resulting BLUE(1-4) scores either approximately equal or surpass the Chen et al performance using the similar meshed-memory transformer. For comparison, the metrics for the state of the art Miura et al. model exceed our model performance; their CheXbert metric is 0.567 and their BLEU4 score is 11.4. The Miura et al. model has the same architecture as our single-feature models but uses reinforcement learning to improve semantic quality and clinical relevance, thus exceeding the metrics achieved in our model experiments as expected. Our models are not trained additionally with reinforcement learning in order to better isolate the effect of pre-training strategy for the visual feature extractor.

CheXbert derived F1 scores by model and pretraining type are included in Figure 3. A common effect is seen across the 14 conditions that for less frequent conditions, there is often a more significant gap between ImageNet model performance and the CheXpert model or double feature model performance. Note that ImageNet model performance does achieve the greatest F1 score of any model on cardiomegaly, atelectasis, and pleural effusion. The double feature model achieves the best performance of all models on edema, consolidation, pneumothorax, support devices, and overall F1 score. The conditions that have high F1 scores commonly also show high performance of the ImageNet model; however, in all these cases the CheXpert model and double feature model also achieve similarly high performance. These results suggest pretraining could be potentially determined by task specificity due to the high variability in the models' performance on varying conditions.

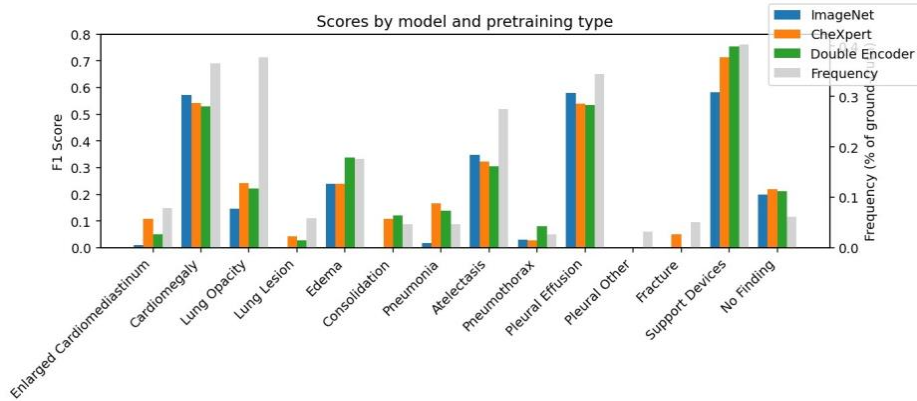

Fig. 3. Clinical CheXbert F1 Scores achieved by the three models and plotted alongside condition frequency in reports

|  | Ground Truth | ImageNet | Double Feature |
| --- | --- | --- | --- |
| Radiograph 1: Negative Findings | Low lung volumes, no pleural effusions. No parenchymal abnormality, in particular no evidence of pneumonia. Borderline size of the cardiac silhouette without pulmonary edema. No hilar or mediastinal abnormalities | In comparison with the study of ___ , there is little overall change. Again there is elevation of the left hemidiaphragm with mild <unk> changes at the left base. No evidence of acute focal pneumonia or vascular congestion. | The lung volumes are low. There is no evidence of pneumonia. No pleural effusions. No pulmonary edema. Normal size of the cardiac silhouette. Normal hilar and mediastinal contours . |
| Radiograph 2: Pacemaker Present | Lungs are fully expanded and clear. No pleural abnormalities. Severe cardiomegaly and cardiomediastinal hilar silhouettes are unchanged. Pacemaker and <unk> leads are unchanged in position. No evidence of displaced rib fracture. | PA and lateral views of the chest provided. Left chest wall pacer device is again seen with leads extending to the region the right atrium and right ventricle. There is a left pleural effusion with associated compressive lower lobe atelectasis. The heart is mildly enlarged. The mediastinal contour is normal. No pneumothorax. Bony structures are intact. | PA and lateral views of the chest provided. Left chest wall <unk> is again noted with lead extending to the region the right ventricle. The heart remains mildly enlarged. The lungs are clear without focal consolidation, large effusion or pneumothorax. No signs of congestion or edema. The mediastinal contour is stable. Bony structures are intact. No free air below the right hemidiaphragm. |
| Radiograph 3: Lung Volume Loss and Plural Effusion | In comparison with the study of ___ , there is continued opacification at the left base most likely reflecting pleural effusion and volume loss in the lower lobe. Mild blunting of the right costophrenic angle persists. No evidence of vascular congestion. Right <unk> catheter remains in place. | As compared to the previous radiograph, the lung volumes have decreased. There is a minimal left pleural effusion, restricted to the left. Subsequent areas of atelectasis at the left lung bases. No evidence of pneumonia. No pulmonary edema. No pneumothorax. | As compared to the previous radiograph, the patient has received a new right internal jugular vein catheter. The course of the catheter is unremarkable, the tip of the catheter projects over the inflow tract of the right atrium. There is no evidence of complications, notably no pneumothorax. The lung volumes have decreased, but the left pleural effusion has decreased. The size of the cardiac silhouette is unchanged. |

Fig. 4. Three chest radiograph reports are included in the ground truth with the results of the ImageNet model and Double Feature Model compared alongside.

## V. DISCUSSION

The results suggest three conclusions: first that ImageNet pretraining provides significantly less overall knowledge for chest radiograph report generation, second that the double feature model is a promising architecture for future medical image captioning tasks, and third that the choice of what to pretrain on is likely more task dependant than previously thought to be within the larger medical domain.

The single feature model pretrained on ImageNet results in lower BLEU(1-4) scores, CIDEr score, and ROUGE-L score than the models using CheXpert pretraining. Additionally, the ImageNet pretrained model results in lower CheXbert clinical F1 score performance than the models using CheXpert pretraining. Scoring lower across all metrics, both those such as BLEU(1-4) that have been shown to be domain agnostic and to value grammatical outputs [28, 29], as well as clinical metrics (CheXbert F1 metric) that score highly the identification of fourteen medical labels, we provide evidence to suggest that the ImageNet pretrained model shows lower performance in report generation solely due to its pretraining on ImageNet. This suggests that ImageNet pretraining does not generalize well to chest radiograph captioning. Figure 3 demonstrates that when ImageNet pretraining results in a higher performing model for a specific condition by CheXbert F1 score, that this is almost always accompanied by the condition itself appearing more frequently in radiographs, demonstrated by cardiomegaly, atelectasis, and pleural effusion. However, for less frequent conditions such as enlarged cardiomediastinum, lung lesion, consolidation, pneumonia, pneumothorax, and fracture, the ImageNet pretrained model is significantly lower in CheXbert F1 score on those conditions than the two models which incorporate CheXpert pretraining. This is likely due to ImageNet pretraining being unable to identify motifs or segment useful areas in rare conditions. Furthermore, ImageNet has been shown in prior work to not generalize well to medical segmentation tasks [15] and thus for developing models intended to be used on images lacking clear morphological boundaries and features, ImageNet pretraining is likely especially a poor choice. However, for conditions with increased morphological information, such as cardiomegaly, ImageNet pretraining for the model may be a suitable choice.

The results of our work suggest that the double feature model is a potential architectural improvement for report generation from medical images. This architecture provides both domain specific information as well as ImageNet morphological and classification benefits. The double feature model achieved the highest BLEU(1-3), ROUGE-L, and clinical CheXbert F1 scores in our experiments. The model's benefit is further realized in the model's performance on conditions such as edema that couple physical morphological features with chest domain specific features; the double feature model achieves the highest CheXbert F1 score on this condition. The application of such models to other medical imaging captioning tasks will rely on the successful development of classification algorithms with image inputs to label outputs for the specific domain. In the chest radiograph space this has been already accomplished by CheXpert models. Recent weakly supervised techniques to harness pseudo labelled data can catalyze development of these models in other medical domains.

Limitations of our approach include that there are other pretraining methods using other datasets that could be studied for potential performance effects for this task. Another such limitation is that applying reinforcement learning could affect pretraining strategy and should be studied in future work.

Below in Figure 4 we include three radiology reports demonstrating the ImageNet pretrained model and double feature model results on radiographs with negative findings, a pacemaker present and cardiomegaly, and lung volume loss/pleural effusion respectively. They are included to contrast the effect of including CheXpert pretraining alongside the pretraining of the ImageNet model to better understand the effect of including CheXpert information.

## VII. Conflict of Interest Statement

Declarations of Interest: None. Sources of Funding: None.

## VIII. Code Availability

The code used for this project is available at https://github.com/evendrow/ifcc.